\documentclass{aa}
\usepackage{txfonts}
\usepackage{graphicx}
\usepackage{setspace}
\usepackage{natbib}
\usepackage{amssymb}
\bibpunct{(}{)}{;}{a}{}{,}

\begin{document}

\newcommand{\fe}{[\ion{Fe}{ii}]}
\newcommand{\Ti}{[\ion{Ti}{ii}]}
\newcommand{\s}{[\ion{S}{ii}]}
\newcommand{\oi}{[\ion{O}{i}]}
\newcommand{\mdot}{$\dot{M}_{jet}$}
\newcommand{\h}{H$_2$}
\newcommand{\kms}{km\,s$^{-1}$}
\newcommand{\um}{$\mu$m}
\newcommand{\lam}{$\lambda$}
\hyphenation{mo-le-cu-lar pre-vious e-vi-den-ce di-ffe-rent pa-ra-me-ters ex-ten-ding a-vai-la-ble co-vers con-ti-nuum re-la-ti-ve clo-sest ve-lo-ci-ty co-rres-pon-ding tem-pe-ra-tu-res res-pon-si-ble me-cha-nism me-cha-nisms ki-ne-ma-tics cha-rac-te-ri-sed pre-vious-ly va-lues theo-re-ti-cal laun-ching }

\title{IR diagnostics of embedded jets: velocity resolved observations of the HH34 and HH1 jets \thanks{Based on observations collected at the European Southern Observatory, La Silla, Chile (ESO programmes 0.74.C-0286(A)).}}
\author{Rebeca Garcia Lopez \inst{1,2,4}\and Brunella Nisini \inst{1} \and Teresa Giannini \inst{1} \and Jochen Eisl\"offel \inst{2} \and Francesca Bacciotti \inst{3} \and Linda Podio  \inst{3}}

\offprints{R. Garcia Lopez, \email{garcia@mporzio.astro.it}}

\institute{INAF-Osservatorio Astronomico di Roma, Via di Frascati 33, I-00040 Monteporzio Catone,
Italy \and Th\"uringer Landessternwarte Tautenburg, Sternwarte 5, D-07778 Tautenburg, Germany \and 
INAF-Osservatorio Astrofisico di Arcetri, Largo E. Fermi 5, I-50125 Florence, Italy \and Universit{\`a} degli Studi di Roma "Tor Vergata" - Dipartimento di Fisica, via della Ricerca Scientifica 1, I-00133 Roma, Italy}
 
%
\date{Received date; Accepted date}
%
%
%
\titlerunning{Velocity resolved NIR spectroscopy }
\authorrunning{Garcia Lopez, R. et al.}

\abstract 
{We present VLT-ISAAC medium resolution spectroscopy of the HH34 and HH1 jets.}
{Our aim is to derive the kinematics and the physical parameters and to study how they vary with jet velocity.}
{We use several diagnostic lines, such as $\fe\,1.644\,\mu m$, $1.600\,\mu m$ and H$_2 2.122\mu m$, in order to probe the atomic and molecular components.}
{In the inner jet region of HH34 we find that the atomic and molecular gas present two components at high and low velocity (the so-called HVC and LVC). The \fe\ LVC in HH34 is detected up to large distances from the source ($>1000$\,AU), at variance with TTauri jets. In H$_2\,2.122\,\mu m$, the LVC and HVC are spatially separated, with an abrupt transition from low to high velocity emission at $\sim1\farcs5$. We moreover detect, for the first time, the fainter red-shifted counterpart down to the central source. In HH1, we trace the jet down to $\sim1\arcsec\,$ from the VLA1 driving source: the kinematics of this inner region is again characterised by the presence of two velocity components, one blue-shifted and one red-shifted with respect to the source LSR velocity. We interpret this double component as arising from the interaction of two different jets. We suggest that the red-shifted component could be part of the HH501 jet. Electron densities and mass fluxes have been measured separately for the different velocity components in the HH34 and HH1 jets. In the inner HH34 jet region, n$_e$ increases with decreasing velocity, with an average value of $\sim1\times 10^{4}$\,cm$^{-3}$ in the HVC and $\sim2.2\times 10^{4}$\,cm$^{-3}$ in the LVC. Up to $\sim10\arcsec\,$ from the driving source, and along the whole HH1 jet an opposite behaviour is observed instead, with n$_e$ increasing with velocity. In both jets the mass flux is carried mainly by the high-velocity gas: lower limits on the mass flux of $3-8\times 10^{-8}$\,M$_{\odot}$yr$^{-1}$ have been found from the luminosity of the $\fe\,1.644\,\mu m$ line. A comparison between the position velocity diagrams and derived electron densities with models for MHD jet launching mechanisms has been performed for HH34. While the kinematical characteristics of the line emission at the jet base can be, at least qualitatively, reproduced by both X-winds and disc-wind models, none of these models can explain the extent of the LVC and the dependence of electron density with velocity that we observe. It is possible that the LVC in HH34 represents denser ambient gas entrained by the high velocity collimated jet.}
{}

\maketitle
%

\section{Introduction}
Protostellar jets and accretion discs are phenomena intimately connected. The relation between both structures and the jet  launching mechanism is, however, still not well understood. Several models have been developed to constrain the physical mechanism by which mass is accelerated from the star's vicinity (\citealt{shu95,ferreira97};\citealt{stellar_wind05}) by the action of so-called ``rotating magnetospheres'' (\citealt{camenzind90}). Observations of the inner part of the  jet structure are required, however, in order to verify the predictions of such  kinds of models.  In particular, NIR spectroscopy is an important tool to investigate the jet structure closest to the driving source. This is especially true for jets in Class 0/I objects, where the high extinction prevents us from  observing the regions close to the source with standard optical tracers.
Several observational studies have indeed been carried out during recent years employing near-IR line diagnostics on jet beams from young embedded sources (e.g. \citealt{nisini_hh1,davisSVS13,linda,takami}). In these studies the main lines investigated are \fe\ lines (e.g. 1.644\,\um\ and 1.600\,\um\,) and \h\ lines (the 2.12\,\um\, line is among the brightest lines). These works have addressed both the physical parameters and the kinematical properties of the sources, and their similarities with more evolved Classical T Tauri Stars (CTTSs). Kinematically,  both Class I and CTT stars, show a  similar behaviour, with the presence of a collimated and large scale jet at high velocity ($\sim$200-400\,\kms\,), called the High Velocity Component (HVC), and a compact gas component localised around the central source at lower velocity ($\sim$0-50\,\kms\,) named Low Velocity Component (LVC).  This supports the assumption that the accretion and ejection in Class I objects occurs by the same mechanism as for T Tauri stars. Synthetic Position Velocity (PV) diagrams constructed for different classes of MHD jet models (i.e. disc-winds, \citealt{garcia_warm} and X-winds, \citealt{shang02}) predict the presence of different velocity components at the jet base, although the details of the observed kinematical features are still not well reproduced by these models. 

On their physical properties, however, jets from Class I and CTTSs show some differences: in general, densities and mass fluxes derived for jets from Class I sources are higher than those in jets from CTTSs, as one would expect in sources with higher accretion rates, embedded in a dense environment (e.g. \citealt{davis03}, \citealt{simone07}). Moreover, the molecular component traced by \h\, NIR lines is significantly enhanced in the younger sources, and in some of them the jet shows up only in molecular form.  

In this framework, an important observational test is to measure the relevant physical parameters separately in the different velocity components of the flows. This kind of diagnostics can show if the differences in the physical properties between jets from Class I sources and CTTSs persist in both the HVC and LVC, and provides further constraints on the origin of these latter. Differences in the excitation conditions of the various velocity components have been measured in 
jets from CTTSs. Such a kind of study has, however, never been attempted for jets from Class 0/I stars.
\begin{table}
\begin{minipage}[t]{\columnwidth}
\caption{List of detected lines}
\label{tab:lambdahh}
\centering
\renewcommand{\footnoterule}{}  
\begin{tabular}{c c c c}
\hline \hline \\[-5pt]
Line id.& $\lambda$\footnote{Vacuum wavelengths in microns.}     &	Jet	  \\

\hline

\fe $a^4D_{3/2}-a^4F_{7/2}$ & 1.5999 & HH34, HH1 \\
\fe $a^4D_{7/2}-a^4F_{9/2}$  & 1.6440 & HH34, HH1\\
\fe  $a^2P_{3/2}-a^4P_{3/2}$  & 2.1333 & HH34, HH1\\
\fe $a^{2}H_{11/2}-a^2G_{9/2}$ & 2.2244 & HH34\\
\Ti  $a^2F_{5/2} - a^4F_{3/2}$ & 2.1605 & HH1\\
\Ti  $a^2F_{7/2} - a^4F_{5/2}$  & 2.0818 & HH1\\
H$_{2}$ 1-0 S(1) & 2.1218 & HH34, HH1\\
H$_2$ 2-1S(2) & 2.1542 & HH1\\
H$_{2}$ 1-0 S(0)  & 2.2235 & HH34\\
H$_{2}$ 2-1 S(1)  & 2.2477 & HH34  \\
Br$\gamma$ & 2.1661 & HH1\\

\hline
\end{tabular}
\end{minipage}
\end{table}


Here, we present recent results on the HH34 and HH1 jets, two classical outflows that are bright in the near-IR (e.g. \citealt{davis00,davis_MHEL,jochen00}). Their driving sources are HH34 IRS and VLA1, two young embedded Class 0/I objects situated at a distance of 450\,pc and 460\,pc, respectively, in a high phase of accretion (\citealt{simone07, chini01}). Velocity resolved, near-IR spectroscopy of HH34, addressing the kinematics of the inner region, has been presented by \citet{davis_MHEL}, \citet{davis03} and \citet{takami}.
More recently, we have  observed the HH34 and HH1 jets in low resolution spectroscopy from 0.6 to 2.4\,\um, and derived the relevant physical parameters of the jets, as a function of the distance from the exciting source, adopting an analysis combining optical and near-IR  line ratios (\citealt{nisini_hh1,linda}). 
\begin{figure}[!ht]
\begin{center}
\resizebox{\hsize}{!}{\includegraphics[totalheight=5cm]{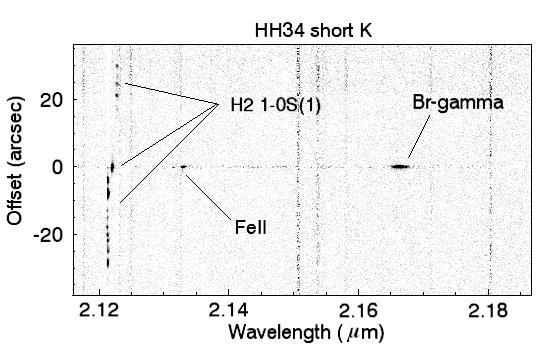}}
\resizebox{\hsize}{!}{\includegraphics[totalheight=8cm]{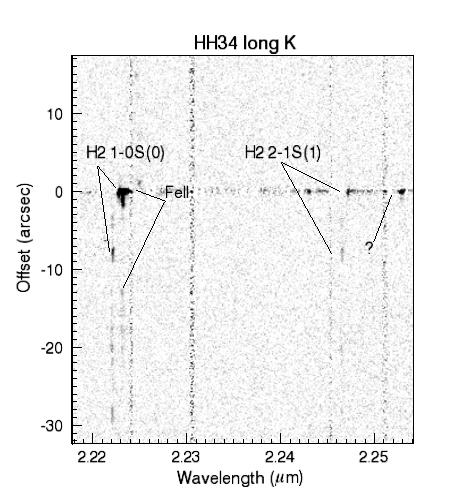}}
\caption{\label{fig:shortK}Continuum-subtracted K-band spectral images of the HH34 jet.}
\end{center}
\end{figure}
In this work, we present velocity-resolved NIR spectra obtained with the spectrograph ISAAC on the Very Large Telescope (VLT) of the European Southern Observatory (ESO). This instrument covers important diagnostic lines of the ionised  (\fe\,1.644, 1.600 $\mu$m) and molecular (\h\, 2.122 $\mu$m) gas. Our aim was studying the kinematics of the atomic and molecular components, and to derive the electron density and mass ejection rates in the different velocity components. The paper is structured as follows:  the observations and data reduction methods are described in Section 2. In Section 3,  we present the results on the kinematics of the \fe\, and H$_2$ lines. In Section 4,  we derive the electron density and mass flux from the \fe\, line luminosity and ratios, and discuss the variation of these parameters as a function of velocity and distance from the central object. In Section 5, we discuss the comparison of our results with the predictions of proposed jet launching models. Finally, we draw our conclusions in Section 6.

\section{Observations and data reduction}

\begin{figure*}[!ht]
\begin{center}
\resizebox{\hsize}{!}{\includegraphics{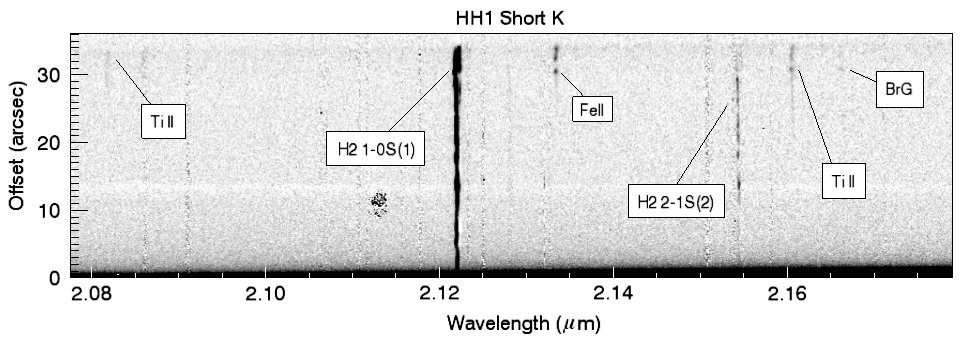}}
\caption{\label{fig:k_hh1}
Continuum-subtracted K-band spectral image of the HH1 jet. }
\end{center}
\end{figure*}


Observations were obtained on 28-29 December 2004 at the ESO VLT telescope on Cerro Paranal, Chile, using the infrared spectrograph and camera ISAAC at medium resolution. The adopted slit width is 0\farcs3\,, corresponding to a nominal resolution R$\sim$8900 in the K-band and 10\,000 in the H-band. The spatial scale of the camera is 0\farcs146/pixel.
We have taken H (1.57-1.65\,\um) and K (2.07-2.20\,\um) band spectral segments with the slit aligned along the jets. An additional segment in K-band, covering the range 2.19-2.31\,\um, was also acquired on HH34. The adopted position angles (PA) were 145\degr\, and -15\degr\, for the HH1 and HH34 jets, respectively. Total integration times were 3300\,s in the K-band for both the jets and 5100\,s and 6900\,s in the H-band for HH1 and HH34, respectively. 
Data reduction was performed using standard IRAF\footnote{IRAF (Image Reduction and Analysis Facility) is distributed by the National Optical Astronomy Observatories, which are operated by AURA, Inc., cooperative agreement with the National Science Foundation.} tasks. The IR spectra were corrected for the atmospheric spectral response by dividing the object spectra by the spectrum of a B5III star showing telluric lines. Wavelength calibration was performed using the atmospheric OH emission lines. The H and K-band effective resolution measured from the OH lines is $\sim$8600 and $\sim$7800 for both objects. Flux calibration was performed using a photometric standard star. The standard stars have been observed at similar air mass and seeing conditions as the objects, therefore no corrections for flux losses due to the slit width smaller than the seeing ($\sim$0\farcs6 in both the nights) have been applied. This procedure probably leads to some flux losses in the more external section of the jet ($\gtrsim$ 15\arcsec\, from the source) where the intrinsic jet diameter, as measured from HST images (\citealt{reipurth_vt_hh34} widens to more than 0\farcs6.
In the case of HH34, we subtracted the continuum emission from the HH34 IRS source using the IRAF task BACKGROUND. 

In the H-band spectral segments, only the \fe\,1.5999 and 1.6440\,\um\, lines have been detected, while several additional transitions have been observed in K-band. Figures \ref{fig:shortK} and \ref{fig:k_hh1} report the K-band spectral images showing the different emission lines found for the HH34 and HH1 jets, while  the most prominent identified lines are listed in Table \ref{tab:lambdahh} for both the jets. For HH34, we report here only the lines which are spatially resolved along the jet direction: additional transitions, mostly from permitted species, have been detected in the on-source spectrum and discussed in \citet{simone07}. Along the HH34 jet, our detected lines are consistent with those observed by \citet{takami} except for our additional detection of the \fe\, 2.224\,\um\, line, originating from the $^2 H$ term. The \fe\, 2.133\,\um\, line, connecting the $^2 P$ and $^4 P$ terms, has been detected in both the HH34 and HH1 jets: as pointed out by \citet{takami}, this transition, with excitation energies in excess of 25\,000 K,  is very useful to probe high excitation regions, such as those found at the jet base or in high velocity shock interaction. In  HH1, we also detected two transitions from \Ti , characterised by excitation temperatures on the order of 7000 K.
Finally, in knot H of HH1, weak emission from Br$\gamma$ is detected, testifying for the high excitation conditions in this knot.


\section{\fe\ and \h\ kinematics}

In order to study the kinematics of the two jets in both the atomic and molecular components, we have 
constructed position-velocity (PV) diagrams of the \fe\ 1.644\,\um\ and \h\ 2.122\,\um\ emission lines. The velocity is expressed with respect to the local standard of rest for both PV diagrams. A parental cloud velocity of 8\,\kms\,\ and 10.6\,\kms\,\ has been adopted for HH34 and HH1, respectively (\citealt{v_cloud_hh34, v_cloud_hh1}), and subtracted in the final PV velocity scale. Distance scales (in arcsec) have been measured with respect to HH34 IRS and VLA1 for HH34 and HH1. Since the driving source VLA1 of the HH1 jet is not visible in the near-IR, we have used the bright Cohen-Schwartz (CS) star as the positional reference, adopting for it an angular separation of 35\farcs9 from the source VLA1. In the following we will describe the results obtained for the two sources separately.

\begin{figure}[!ht]
\begin{center}
\includegraphics[totalheight=13.7cm]{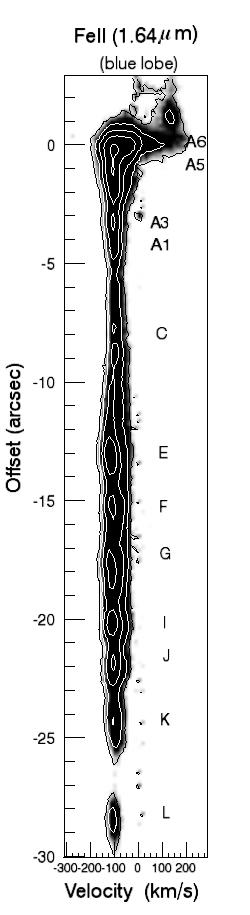}
\includegraphics[totalheight=13.7cm]{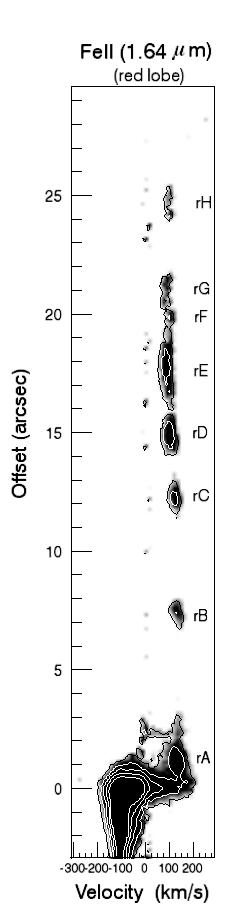}
\caption{\label{fig:largescaleFe}
Continuum-subtracted PV diagrams of the \fe\ 1.644\,\um\ emission line for the blue and red lobe of the HH34 jet. A P.A. of -15\degr\, was adopted. Contours show values of 5, 15, 45, 135, 405\,$\sigma$ for 
the blue lobe, and 5, 10, 20, 40, 80\,$\sigma$ for the red lobe. On the Y-axis the distance from HH34 IRS is reported.}
\end{center}
\end{figure}
\begin{figure}
\begin{center}
\includegraphics[totalheight=13.5cm]{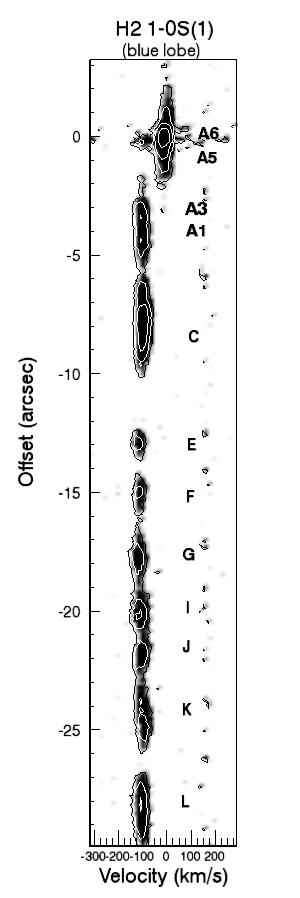}
\includegraphics[totalheight=13.5cm]{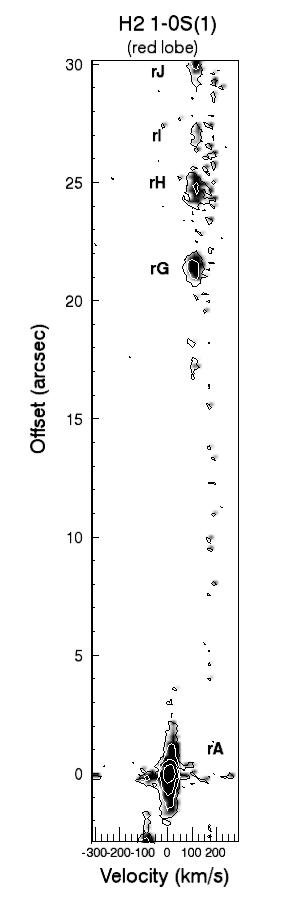}
\caption{\label{fig:largescaleH2}
Continuum-subtracted PV diagram for the \h\ 1-0S(1) emission line for the blue and red lobe of the HH34 jet. A P.A. of -15\degr\, was adopted. Contours show values of 5, 15, 45, 135, 405\,$\sigma$ for both
the lobes. On the Y-axis the distance from HH34 IRS is reported.}
\end{center}
\end{figure}
\subsection{HH34}
\subsubsection{Large scale properties}

Figure \ref{fig:largescaleFe} and \ref{fig:largescaleH2} show  the \fe\ 1.644\,\um\ and \h\ 2.122\,\um\ lines PV diagrams relative to the blue-shifted and red-shifted lobes of HH34. The intensity scale is different in the two figures to evidence the relatively weak red-shifted emission. The redshifted counterpart of HH34 jet is clearly detected  in our spectral images. We have named the detected blue-shifted emission knots from A to L, following the nomenclature of \citet{jochen_hh34} and \cite{reipurth_vt_hh34}. We do not detect emission, however, from the knot B, since this knot is not aligned with the main jet axis, as evidenced in \cite{reipurth_vt_hh34}. The next knots we observe are the knots C and D that we have group together as knot C. The red-shifted knots, which are approximately located at symmetric positions with respect to the corresponding blue-shifted knot, are named here from rA to rH.

\begin{table*}
\begin{minipage}[t]{\textwidth}
\caption{Observed radial velocities along the HH34 jet.}
\label{tab:v}
\centering
\renewcommand{\footnoterule}{}  
\begin{tabular}{cc|cc|cc|cc|c|cc}

\hline \hline \\[-5pt]

   \multicolumn{6}{c}{Blue lobe} &
 \multicolumn{5}{c}{Red lobe}\\
  \hline\\[-5pt]                                 
 &  & \multicolumn{2}{c}{\fe\ 1.64 $\mu$m}&
       \multicolumn{2}{c}{\h\ 2.12 $\mu$m}&  
 &
 &
\fe\ 1.64 $\mu$m &
\multicolumn{2}{c}{\h\ 2.12 $\mu$m} \\
Knot &
r$_t$\footnote{Distance from the source in arcsec given by the mean value in the adopted aperture. Negative values 
correspond to the southeastern, blue-shifted jet axis.}	&
HVC\footnote{Radial velocities (in\,\kms\,) with respect to the local standard 
of rest and corrected for a cloud velocity of 8\,\kms\,. The radial velocity 
error is 2\,\kms\,. LVC and HVC refer to the  Low Velocity Component and High 
Velocity Component. The velocity dispersion (in \kms\,), measured from the line FWHM deconvolved for the instrumental profile, is reported in brackets for the HVC.}	&	
LVC$^b$	&	
HVC$^b$	&
LVC$^b$	& 
Knot &
r$_t^a$	&
 &
HVC$^b$	&
LVC$^b$ \\
\hline\\
A6  & 0.0  & -92  (46) & -52 & -89 (30) & -4  & rA & +1.0 & +133 (43) &	    & +10   \\
A3  & -3.0 & -98  (37) & -67 & -98 (17) & -15 & rB & +7.3 & +141 (51) &	    &       \\
A1  & -4.5 & -93  (37) & -68 & -95 (12) & -7  & rC & +12.3& +128 (58) &	    &	   \\
C   & -9.0 & -92  (43) &     & -94 (12) & -7  & rD & +15.0& +108 (35) &	    &	      \\
E   & -12.5& -108 (44)	&     &	-110 (8) &     & rE & +17.4& +96 (43)	&	    &	       \\	
F   & -15.2& -100 (43)	&     &	-107 (12)&     & rFG& +21.0& +96 (46)	& +113 (26)  &	             \\
G   & -17.2& -106 (43)	&     &	-111 (8) &     & rH & +25.4& +100 (35)	& +115 (21)  & 	\\
I   & -20.0& -100 (43) &     &	-105 (24)&     &    & 		    &		&           &	\\
J   & -22.1& -96  (43) &     &	-99 (19) &     &    &	 	    &		&   	    &		\\
K   & -24.8& -92  (44) &     &	-92 (26) &     &    &	 	    &		&	    &	 \\
L   & -28.9& -98  (26) &     &	-99 (19) &     &    &	 	    &		&	    &	\\
\hline
\end{tabular}
\end{minipage}
\end{table*}


The values of the \fe\ and \h\ 1-0S(1) peak velocities have been computed applying a Gaussian fit to the line profile of every blue and red-shifted knot emission, and they are listed in Table \ref{tab:v}. In the knots closer to the star, where two velocity components have been identified, their peak velocity has been separately measured considering a two-Gaussian fit. The \fe\ radial velocities in the blue lobe cover a range from -92 to -108\,\kms\, , which is consistent with the values measured by \cite{takami} and \cite{davis03}, who found a range in radial velocities for the blue lobe from -90 to -100\,\kms\,, corrected for a cloud velocity of 8\,\kms\,. 

From knots A6 to I the blueshifted radial velocity increases from $\sim$-92 to $\sim$-100\,\kms\, , passing through a maximum at $\sim-108$\,\kms\,, then decreases again down to $\sim$-92\,\kms\,\ at knot K. Finally it increases in knot L to $\sim$-98\,\kms\, . Errors in relative velocities are estimated to be on the order of 2\,\kms\,\,, since our wavelength calibration has an uncertainty of 0.1\AA{}. The \fe\, radial velocity along the red lobe shows a similar behaviour. The velocity roughly increases from the knot closest to the source, rA, to the knot rB, from $\sim$130\,\kms\,\ to $\sim$140\,\kms\,, then decreases down to a value of $\sim$96\,\kms\,. 
Table 2 reports the velocity dispersion of the \fe\, emission, measured from the FWHM of the Gaussian fit, deconvolved for the instrumental profile. In knots A6-A3-A1, where different velocity components are evident (see Section 3.1.2) the reported velocity dispersion refers to the brightest component at high velocity. Intrinsic line widths on the order of 35-40\kms\, are observed all along the jet.
Under the assumption that the line emission arises from unresolved shock working surfaces, we have, following  \cite{hartigan87}, that the shock velocity is roughly given by $V_s \sim \Delta V$(FWZI)$\sim 2\times \Delta V$(FWHM). This implies shock velocities of the order of 70-80\kms, thus much higher than the value of $\sim$30\kms\, estimated by \cite{hartigan94}  on the basis of the comparison of optical line ratios with shock models. Indeed, shocks with speeds as high as 80\kms\, are expected to produce a strong ionisation, on the order of $x_e\sim 0.3-0.4$ (\citealt{hartigan94}) while \cite{linda} measured an average ionisation along the HH34 jet of only 0.04. Therefore, it seems that the line widening is  determined not only by the shock but also by, e.g. a lateral expansion of the jet. 

\h\ radial velocities have a range from -89 to -110\,\kms\, for the blue lobe and an average value of $\sim$+115\,\kms\, for the red lobe. The \h\, line profile is resolved all along the jet, with $\Delta V$ values of the order of 10-30 \kms\,.

Also in \h\ , radial velocities show cyclic variations on small scales along the jet, with an increase of roughly 20\,\kms\, from knot A6 to knot E and a subsequent decrease of the same order at knot K.
\begin{figure}[!ht]
\begin{center}
\includegraphics[totalheight=7cm]{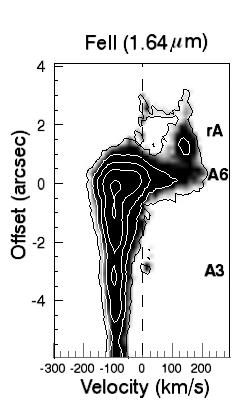}
\includegraphics[totalheight=7cm]{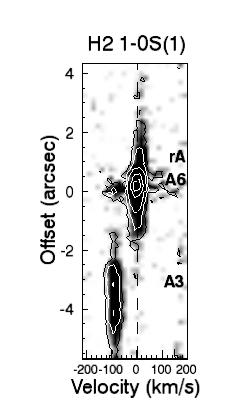}
\caption{\label{fig:shortscale}
Continuum-subtracted PV diagrams for the \fe\ 1.644\,$\mu$m and \h\ 1-0S(1) emission lines of the HH34 jet in the region nearest to the source. A P.A. of -15\degr\, was adopted. Contours show values of 5, 15, 45, 135, 260\,$\sigma$ for both lines. On the Y-axis the distance from HH34 IRS is reported.}
\end{center}
\end{figure}
It is known that the HH34 jet presents velocity variability on large and small spatial scales. \citet{raga98} and \citet{raga02} have shown that to reproduce the velocity pattern observed at different epochs, a model of variable ejection velocity including three modes with different periods is needed. The fastest of these modes can be represented with a sinusoid having a period of 27 yrs and an amplitude of $\sim$15\,\kms\,. This is roughly consistent with our observed velocity pattern, that is reproduced quite closely by both \fe\ and \h\ and by the red-shifted and blue-shifted gas, clearly indicating its origin from ejection velocity variability.
The velocity pattern observed at large distance may, however, be biased by the change in the jet axis direction that occurs at d$\sim$5\arcsec\, from the source, coupled with the widening of the jet diameter (up to $\sim$0\farcs6 at $\sim$20 arcsec from the source, \citealt{reipurth_vt_hh34}). Adopting an instrumental slit of only 0\farcs3, part of the kinematical components of the jet may not be properly probed by our observations.
\begin{figure}[!ht]
\begin{center}
\includegraphics[totalheight=13.8cm]{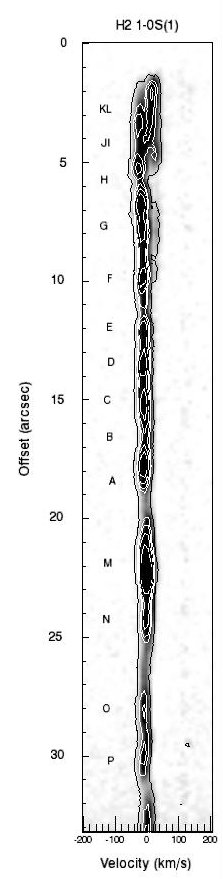}
\caption{\label{fig:hh1_212}
Continuum-subtracted PV diagram for the H$_2$ 1-0 S(1) line along the HH1 jet. A position angle of 145\degr\, was adopted. Contours show values of 10, 40, 60, 100\,$\sigma$. On the Y-axis the distance from the driving source VLA1 is reported. }
\end{center}
\end{figure}

\subsubsection{Small scale properties}

Close to the central source, the \fe\ lines broaden and emission at lower velocities, down to 0\,\kms\, appears within $\sim$3\arcsec\, from the central source (Figure \ref{fig:shortscale}). Inside $\sim$1\arcsec\, we see emission also at positive velocities, reconnecting with the spatially resolved red-shifted knot rA.
This central region was already observed in \fe\ and \h\ by \citet{davis_MHEL,davis03} and \cite{takami}. Our observations have, however, a better spatial resolution than \cite{davis_MHEL,davis03} and are much deeper than those of \cite{takami}, who did not detect the \fe\ redshifted component that we see at the jet base.  
We identify the high blue-shifted velocity corresponding to the large scale jet as the High Velocity Component (HVC) and the emission component from 0 to $\sim$50\,\kms\, the Low Velocity Component (LVC), in analogy with the HVC and LVC observed in the forbidden emission line (FEL) regions of T Tauri stars. We remark, however, that in T Tauri FELs one usually observes two separated peaks spatially located at different offsets from the central source, the HVC being commonly displaced further downstream (e.g. \citealt{hirth97}). Here the HVC has a peak on-source while the LVC is seen as a weaker shoulder of the brightest component. This behaviour could, however, be due to the lower resolution of our observations as a consequence of the larger distance of HH34 with respect to other studied T Tauri stars.
When compared with other \fe\ PV diagrams observed in T Tauri stars, HH34 is more similar to HL Tau and RW Aurigae \citep{pyo06} than to L1551 and DG Tau \citep{pyo02,pyo03}.

At variance with \fe, the H$_2$ PV diagram shows spatially and kinematically separated LVC and HVC, and only the LVC is visible down to the central source. This component is close to 0\,\kms\, LSR velocity and the blue-shifted and red-shifted jets differ by less than 10\,\kms\,. The HVC appears at a distance of 2\arcsec\ from the central source, at the position of the knots A6 and rA. At intermediate velocities between these two components, no emission is seen in the H$_2$ PV diagram. This suggests that the two components correspond to physically distinct regions. The origin of the H$_2$ high and low VC will be further discussed in Section 5.1.

\subsection{The HH1 jet}
%
\begin{table}
\begin{minipage}[t]{\columnwidth}
\caption{Observed radial velocities along the HH1 jet.}
\label{tab:vhh1}
\centering
\renewcommand{\footnoterule}{}  
\begin{tabular}{cc|c|cc}
\hline \hline\\[-5pt]
Knot & r$_t$\footnote{Distance from the source in arcsec given by the mean value in the adopted aperture.} &\fe\ 1.64 
$\mu$m & \multicolumn{2}{c}{H$_2$ 2.12 $\mu$m} \\ 
     & (\arcsec) &    V$_r$\footnote{Radial velocities (in\,\kms\,) with 
respect to the LSR and corrected for a cloud velocity of 10.6\,\kms\,. The 
radial velocity error is 2\,\kms\,. The velocity dispersion (in \kms\,), measured from the line FWHM deconvolved for the instrumental profile, is reported in brackets for the HVC.} 
    & \multicolumn{2}{c}{V$_r^b$} \\
 &  &  & Blue & Red \\
\hline \\[-5pt]
KL & 1.6 & -28 (64) &-16  & +13  \\
JI & 3.4 & -46 (53) &-27  & +6  \\
H  & 5.1 & -43 (47) &-24  & +14  \\
G  & 7.5 & -35 (22) &-17  & +23  \\
F  & 9.8 & -31 (20) &-14  & +17    \\
E & 11.6 & -24 (29) &-11  &  \\
D & 13.6 &	     &-12  &  \\
C & 15.0 &	     &-10  &  \\
B & 17.0 &	     &-11  &  \\
A & 18.2 &	     &-3   &  \\
M & 22.1 &	     &-4   &  \\
N & 23.5 &	     &-12  &  \\
\hline 
\end{tabular}
\end{minipage}
\end{table}

\begin{figure*}[!ht]
\begin{center}
\includegraphics[totalheight=8cm]{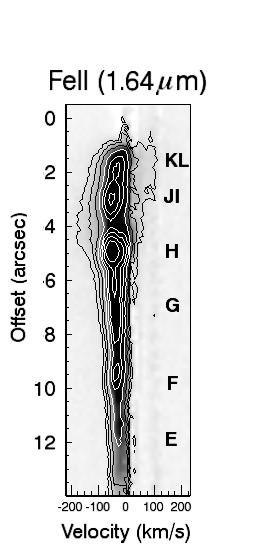}
\includegraphics[totalheight=8cm]{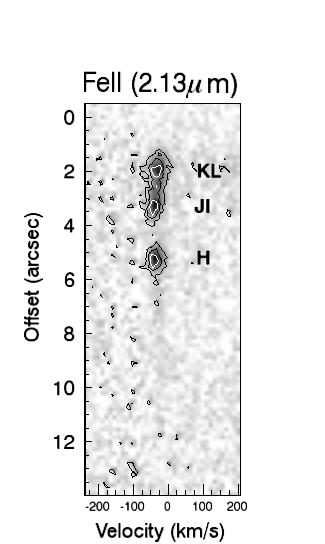}
\includegraphics[totalheight=8cm]{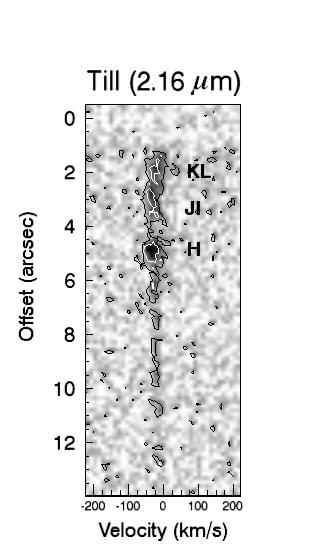}
\caption{\label{fig:hh1_164}
PV diagrams for the \fe\ 1.64\,$\mu$m, 2.13\,$\mu$m and \Ti\ 2.16\,$\mu$m emission lines of the HH1 jet. A position angle of 145\degr\, was adopted. Contours show values of 11, 22,...,176\,$\sigma$; 4, 8, 12\,$\sigma$ and 3, 6, 12\,$\sigma$, respectively. On the Y-axis the distance from the driving source VLA1 is reported.}
\end{center}
\end{figure*}

Figures \ref{fig:hh1_212} and \ref{fig:hh1_164} show the PV diagrams of the H$_2$\,2.122\,$\mu m$ and \fe\,1.644\,$\mu m$  lines of the HH1 jet. The PVs of the  \fe\,2.133\,$\mu m$ and \Ti\,2.160\,$\mu m$ lines are also presented in Figure \ref{fig:hh1_164}. The observed knots are named from KL to P following the nomenclature by \cite{jochen94} and \cite{bally}. \fe\ is detected only in the knots closer to the star, from KL to E, while H$_2$ can be traced all along the jet; as already shown by \cite{davis00} and \cite{nisini_hh1}, the ratio \fe\ /H$_2$ sharply decreases with the distance from VLA1. Such a decrease is accompanied by a decrease of the \h\ and \fe\  radial velocities. 
Also the velocity dispersion diminishes from the internal to the external knots, as shown in Table \ref{tab:vhh1}. In particular, we measure intrinsic FWHM (i.e. deconvolved by the instrumental width) decreasing from  $\sim$ 64\,\kms\, in knots KL, to $\sim$ 20\,\kms\, in knots G-F. As in HH34, these values imply high shock speeds (from $\sim$  120\,\kms\ in knot KL to $\sim$ 40\,\kms\, in knot G), progressively decreasing outwards. Indications of shock velocities in the HH1 jet larger than in HH34 for the internal knots are given by the detection of the Br$\gamma$ and \fe\,\,2.133\,\um\, lines in knot H. These lines are not detected in knots KL and JI probably due to the larger extinction.
The \fe\  and \h\ radial velocities appear double-peaked from knot KL to H in \fe\ and from knot KL to F in \h\,. Both, \fe\ and \h\ present one of the two components red-shifted with respect to the LSR. 

The \fe\ blue-shifted radial velocity increases from knot KL to knot JI from a value of $\sim$-28\,\kms\, to $\sim$-46\,\kms\, and then decreases again to a velocity of $\sim$-24\,\kms\, in knot E. The red-shifted component is, however, too faint to be fitted with a Gaussian profile. Nevertheless, we can estimate an average value of $\sim$+30\,\kms\, from knot KL to H.

\begin{figure}[!ht]
\begin{center}
\includegraphics[totalheight=10cm]{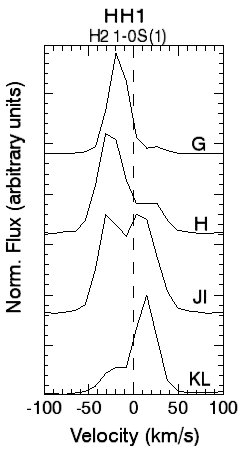}
\caption{\label{fig:intknots_212}
Normalised line profiles of the H$_2$ 1-0S(1) for different knots near the source. }
\end{center}
\end{figure}

On the other hand, in the H$_2$ blue-shifted component the velocity increases from knot KL to knot JI, as in the case of \fe\ and then roughly decreases to $\sim$-3\,\kms\, in knot A. After that, the radial velocity increases again to a value of around $-12$\,\kms\, in knot N. The red-shifted component covers a velocity range from +6 to +23\,\kms\,. The H$_2$ line profiles are represented in Figure \ref{fig:intknots_212} for several knots in the region closest to the source. The measured \h\, velocities are consistent with the ones reported in \cite{davis00} for the knots F and A. The \h\, lines are non-resolved implying an intrinsic line width less than 39\,\kms\ .
The origin of the double velocity component can be due either to the presence of another jet (e.g. the one driving the HH501 object \citealt{Reipurth_hh1,bally}) or to the emission along the wings of a bow shock seen almost perpendicular to the line of sight. These two possibilities will be further discussed in Section 4.1.2.

\section{Diagnostics of physical parameters}

\subsection{Electron density}
\begin{figure*}[!ht]
\begin{center}
\includegraphics[totalheight=6cm]{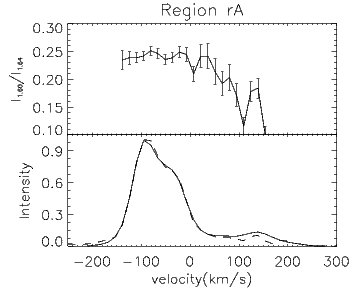}
\includegraphics[totalheight=6cm]{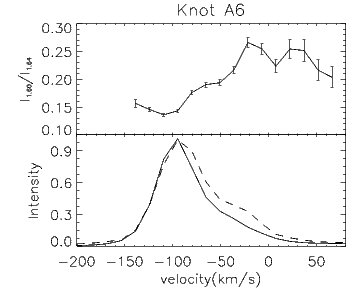}
\includegraphics[totalheight=6cm]{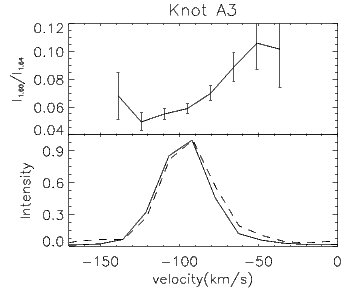}
\includegraphics[totalheight=6cm]{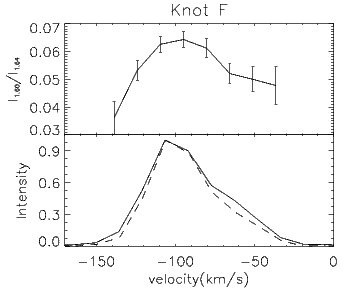}

\caption{\label{fig:ratio_hh34}
Normalised line profiles (lower panels) of the \fe\, lines 1.644\,$\mu$m (solid line) and 1.600\,$\mu$m (dotted line), and their ratio in each velocity channel (upper panel) for different extracted knots along the HH34 jet. Note that rA is a region that includes the knot rA.}
\end{center}
\end{figure*}

\begin{figure*}[!ht]
\begin{center}
\includegraphics[totalheight=6cm]{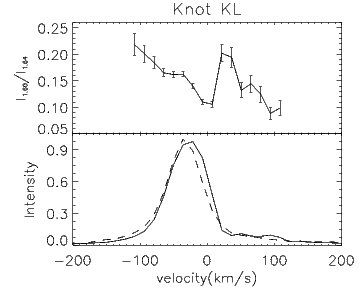}   
\includegraphics[totalheight=6cm]{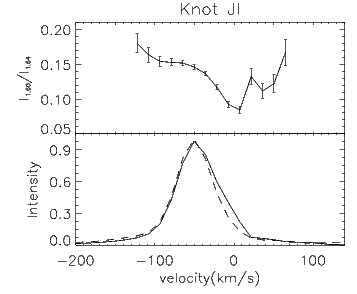}    
\includegraphics[totalheight=6cm]{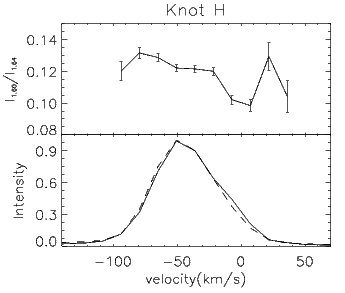}     
\includegraphics[totalheight=6cm]{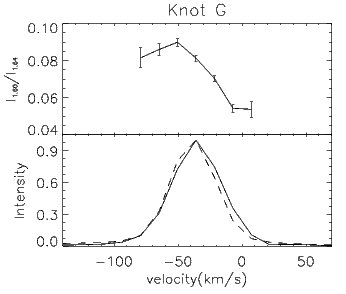}     
\caption{\label{fig:ratio_hh1}
Normalised line profiles (lower panels) of the \fe\, lines 1.644\,$\mu$m (solid line) and 1.600\,$\mu$m (dotted line), and their ratio in each velocity channel (upper panel) for different extracted knots along the HH1 jet.}
\end{center}
\end{figure*}

The electron density in the atomic jet component can be derived from the ratio of the \fe\ 1.600/1.644 $\mu$m lines. This ratio is sensitive to $n_e$ values between $\sim$10$^{3}$ and 10$^{5}$ cm$^{-3}$, while it depends only weakly on the temperature (e.g. \citealt{nisini02}). Electron densities of the \fe\ emission line region as a function of the distance from the driving source has been measured by \cite{linda} and \cite{nisini_hh1} for the HH34 and HH1 jets. In addition, \cite{takami} provide the electron density in the inner region of HH34.
Taking advantage of the velocity resolved profiles in both the 1.644 and 1.600 $\mu$m lines, we can now  measure the electron density in the different velocity components. To do that, we have extracted the spectra of the two lines at different positions along the flows and measured the \mbox{1.600 $\mu$m/1.644 $\mu$m} intensity ratio in each pixel along the spectral profile. 
Figures \ref{fig:ratio_hh34} and \ref{fig:ratio_hh1} show the normalised profiles of the two lines and their ratio as a function of velocity for HH34 and HH1. The spatial intervals used to extract the spectra of the individual knots are given in Tables \ref{tab:mfluxhh34} and \ref{tab:mfluxhh1}. The \fe\, line ratio has been computed only for the velocity points where the intensity in both the lines has been measured with a S/N larger than three. The plotted ratio gives a qualitative indication on  how $n_e$ varies in the different velocity components, that is, a higher ratio indicates a higher electron density.
In the internal knots of HH34 (from A6 to A3)  the 1.600 $\mu$m/1.644 $\mu$m ratio decreases by $\sim$70\% going from $\sim$-50\,\kms\, to -100\,\kms\,. In the knots at larger distance from the central source, on the other hand, the maximum 1.600 $\mu$m/1.644 $\mu$m ratio is observed at the radial velocity peak, with some evidence that the ratio decreases in the line wings at both higher and lower velocities.
In HH1, the 1.600 $\mu$m line has been detected with a S/N larger than three only in the internal knots, from JI to  F. Here, the 1.600 $\mu$m/1.644 $\mu$m ratio has a minimum close to 0\,\kms\, velocity and increases, up to a factor of two in knot KL, towards high velocities. This behaviour is therefore different from what has been observed in the internal knots of HH34. In HH1, however, the inner jet region is not detected due to the high extinction and therefore we are not observing here regions at the jet base as in HH34. It seems indeed that the dependence of the electron density with the velocity is different between the FEL regions close to the star and the knots along the jet beam. 

In order to have a quantitative determination of the electron density in the different velocity components, we have separated in the line profile the contribution from the HV and the LV components, and applied a 16 level Fe$^{+}$ statistical equilibrium model (\citealt{nisini02}). In the following, the results are discussed separately for the two sources.

\subsubsection{The HH34 jet}
In HH34, we have defined the HVC and LVC velocity ranges from the profile of the 1.600 $\mu$m line of knot A6, where the two components have been fitted with a two-Gaussian fit.  We have then measured the line ratios considering the intensities integrated in the FWHM ranges of these two Gaussians also for all the other knots. The considered velocity bins are from $\sim$-120\,\kms\, to $\sim$-66\,\kms\, (HV) and from \mbox{$\sim$-66\,\kms\,} to \mbox{$\sim$-7\,\kms\,} (LV). 
The adopted $A_v$ and $T_e$ values are given in Table \ref{tab:mfluxhh34}, together with the derived electron densities in the two components. Figure \ref{fig:mdot} plots the derived values of n$_e$ as a function of the distance from the source. The first trend that we notice is a sharp decrease of the HVC electron density (from $\sim$10$^4$ to $\sim 2\times 10^3$ cm$^{-3}$) from the knot A6 to the other knots at distances farther than 2\farcs5 from the source. Such a decrease in $n_e$ has also been observed in \cite{linda} and the values they derived, scaled for the different considered spatial regions observed through slits of slightly different width, are consistent with our measured values for $n_e$. In the red-shifted knot rA, which is the only knot where significant 1.600 $\mu$m emission has been detected in the red-lobe, we find a value for $n_e\sim1.2\times 10^4$ cm$^{-3}$, i.e. comparable to the value derived in knot A6. Secondly, as shown in Figure \ref{fig:mdot}, the values for $n_e$ in the LVC are higher  than in the HVC: in knot A6, the LVC electron density is $2.2\times 10^4$ cm$^{-3}$, i.e. 70\% higher than in the HVC. About the same percentage is measured in knots A1 and A3. From the information in our data, we are unable to disentangle, whether the larger electron density in the LVC with respect to the HVC is due to a higher total density or to a higher ionisation fraction.
Our result is, however, in agreement with what is found in the FEL regions of T Tauri stars, where the emission component associated with the low velocity gas is denser and less excited than the HVC (\citealt{hamann94,hartigan95}). In such studies, the LVC was not spatially resolved and the different values for $n_e$ found between the HVC and LVC were interpreted assuming that the LVC originates from a dense compact region close to the disc surface, while the HVC is associated with a more extended high velocity jet displaced further out. Such an interpretation was supported by the spatial offset from the central source often found between the two components of the forbidden lines detected in spatially unresolved spectra of T Tauri stars.
Spatially resolved measurements of the electron density in the HV and LV components have been performed in DG Tau by \cite{bacciotti00} and  \cite{lavalley00}. In this case, at variance with HH34, the electron density has been found to increase with velocity, up to a distance of $\sim$ 450 AU from the central source. On the other hand, in the DG Tau micro-jet the total density in the LV component is higher than in the HV component, because of the much lower ionisation fraction. Therefore, it might be that the density structure in HH34 and in the T Tauri stars is similar, but they differ in the excitation conditions.
In HH34, we spatially resolve the LVC, whose emission is traceable up to $\sim$5 arcsec from the source, i.e. more than 2000 AU. Therefore, the spatial scale between the DG Tau micro-jet and the HH34 jet are very different: in addition, and at variance with many T Tauri stars, we do not observe a spatial offset between the HV and LV components, although this may result from the moderate spatial resolution of our observations. Therefore, it may be that the LVC that we are tracing with our observations has a different origin than in the CTTS case.
In the outer knots, i.e. from C outwards, we cannot distinguish High and Low velocity components anymore. Instead, the density here seems to have an opposite behaviour, with the higher density at the velocity peak and the lower densities in the line wings. The analysis performed in these knots located far from the source could, however, be affected by  a not-perfect alignment of the slit with the jet axis and by the intrinsic jet width larger than the slit, as discussed in Sect. 3.1.1 . Nevertheless, such a pattern agrees with the results obtained by \cite{beck07} using integral field spectroscopy of this part of the jet. They found that both the velocity and the electron density decrease with distance from the jet axis. Such a behaviour is consistent with models for jet internal working surfaces.
\subsubsection{The HH1 jet}
\begin{table*}
\begin{minipage}[t]{\textwidth}
\caption{\mdot\ along the HH34 jet.}
\label{tab:mfluxhh34}
\centering
\renewcommand{\footnoterule}{}  
\begin{tabular}{c|cc|ccc|ccc}
\hline \hline \\[-5pt]
Knot  & r$_t$ \footnote{Distance from the source in arcsec. Negative values correspond to the southeastern, blue-shifted jet axis.} &   A$_{V}$\footnote{Visual extinction from \cite{linda}.} & V$_t$(HVC)\footnote{Tangential velocity assuming an inclination of the jet i=22\fdg7 to the plane of the sky (\citealt{jochen_hh34}).} & n$_e$(HVC)\footnote{Electron density for the HV and LV components.} & $\dot{M}_{jet}$(HVC)\footnote{$\dot{M}_{jet}$ for the HV and LV components assuming an electron temperature of 7000~K.}     & V$_t$(LVC)$^{c}$ & n$_e$(LVC)$^{d}$ & $\dot{M}_{jet}$(LVC)$^{e}$  \\  
  & (\arcsec\,)& mag &	(\kms\ ) &(10$^3$ cm$^{-3}$) &(M$_\odot$ yr$^{-1}$)& (\kms\ )	 & (10$^3$ cm$^{-3}$) &(M$_\odot$ yr$^{-1}$) \\ 

\hline \\[-5pt]
A6&(-2.6,+0.9)  &7.1	&  218	 & 10.5	&$5.2\times 10^{-8}$&	124	&	22.5	&$6.6\times 10^{-9}$ \\
A3&(-3.8,-2.6)  &7.1	&  232	 & 1.8 	&$8.3\times 10^{-8}$&	159	&	5.0	&$2.0\times 10^{-9}$ \\
A1&(-5.4,-3.8)  &7.1	&  220	 & 1.8	&$3.3\times 10^{-8}$&	162	&	5.0	&$1.3\times 10^{-9}$  \\
C &(-9.9,-8.0)  &7.1	&  218	 & 1.2	&$4.4\times 10^{-8}$&		&	    	& 			\\
E &(-14.1,-11.0)&1.3	&  255	 & 1.8	&$5.9\times 10^{-8}$&		&		&  			\\
F &(-16.6,-14.1)&1.3	&  236	 & 1.8	&$5.0\times 10^{-8}$&		&		&		\\ 
G &(-17.8,-16.6)&1.3	&  251	 & 2.5 	&$1.0\times 10^{-7}$&	 	&		&			\\
I &(-21.0,-19.2)&1.3	&  236	 & 2.5 	&$4.2\times 10^{-8}$&		&		&  			\\
J &(-23.3,-21.0)&1.3	&  227	 & 1.2 	&$3.9\times 10^{-8}$&		&		&		  \\
K &(-26.1,-23.5)&1.3	&  217	 & 1.5	&$1.0\times 10^{-8}$&		&		&		\\
L &(-30.3,-27.5)&	&  232	 &    	&	     &	 	&	 	&	 	 \\
rA&(0, +2.0)    &7.1	&  218	 & 12.0	&$2.2\times 10^{-9}$&		&		&		\\

\hline
\end{tabular}
\end{minipage}
\end{table*}

In HH1 we define a low and high velocity component from the \fe\ 1.644 $\mu$m line profile of knot KL. We selected a HVC from $\sim$-80 to $\sim$-22\,\kms\, and a LVC from $\sim$-22 to $\sim$36\,\kms\,. In analogy to HH34, $n_e$ for the LVC and the HVC has been computed separately for all extracted knots. Visual extinction and electron temperature values in each knot have been taken from \cite{nisini_hh1}. At variance with the HH34 jet, the $n_e$ in the LVC is lower than in the HVC. In the HVC $n_e$ decreases knot by knot from a value of $10^4$ cm$^{-3}$ in knot KL to $3.8\times 10^3$\,cm$^{-3}$ in knot F. On the other hand, in the LVC $n_e$ decreases from $9.8\times 10^3$\,cm$^{-3}$ to $8.3\times 10^3$\,cm$^{-3}$ in knot KL and JI, respectively. These $n_e$-values are in agreement with those derived by \cite{nisini_hh1} in velocity integrated spectra. 
 
The presence of two velocity components and the corresponding  dependence on electron density with the  velocity, was already found by \cite{solf91}, who produced a PV diagram of the electron density in the HH1 jet using the optical \s\,$\lambda$\,6716/6731 line ratio. They found a blue-shifted component with an electron density of $\sim$4000\,cm$^{-3}$ and a slightly red-shifted component having a lower density of $\sim$1000-2000\,cm$^{-3}$. They interpreted  the blue-shifted high-density component as due to scattered light originating from a jet region closer to the star, where the density and excitation are higher. This interpretation is, however, difficult to keep in the light of our IR observations. Indeed, the blue-shifted line component is clearly identifiable as the main jet component, extending at large distances from the exciting source. Moreover, our H$_2$ PV diagram (Figure \ref{fig:hh1_212}) clearly shows the presence of two separate components that cannot be attributed to scattered light contribution. The redshifted component peaks at the KL position and decreases in intensity further out, while the second blue-shifted component gradually increases its intensity with the distance.  One possible interpretation is that the red-shifted component belongs to a different jet that intersects the HH1 jet at the KL position. Such a jet could be responsible for the two bright knots designated as HH501 objects by \cite{Reipurth_hh1}, and that are moving away from VLA1 with a proper motion vector inclined with respect to that of the HH1 jet by $\sim$10\degr. If we take the axis of this separate jet equal to the direction of the proper motion vector, the HH501 jet should intersect the HH1 jet at the position of the KL knots. 
\subsection{Mass flux}
\begin{table*}
\begin{minipage}[t]{\textwidth}
\caption{\mdot\ along the HH1 jet.}
\label{tab:mfluxhh1}
\centering
\renewcommand{\footnoterule}{}  
\begin{tabular}{c|ccc|ccc|ccc}
\hline \hline \\[-5pt]
                                    
Knot  & r$_{t}$\footnote{Distance from VLA1 in arcsec.} & A$_{v}$\footnote{Visual extinction and electron temperature from \cite{nisini_hh1}.} & T$_e^{b}$ & V$_{t}$(HVC)\footnote{Tangential velocity assuming an inclination of the jet i=10\degr\, to the plane of the sky (\citealt{bally}).} & n$_{e}$(HVC)\footnote{Electron density for the HV and LV components.} & $\dot{M}_{jet}$(HVC)\footnote{$\dot{M}_{jet}$ for the HV and LV components.} & V$_{t}$(LVC)$^{c}$ & n$_{e}$(LVC))$^{d}$ & $\dot{M}_{jet}$(LVC)$^{e}$ \\
 & (\arcsec)  & mag & ($10^3$ K) & (\kms\ ) & ($10^3$ cm$^{-3}$) &( M$_{\odot}$ yr$^{-1}$)& (\kms\ )  & ($10^3$ cm$^{-3}$) & (M$_{\odot}$ yr$^{-1}$) \\
\hline \\[-5pt]
KL &(0.7,2.6)	&	 8.3	&	11	&	159	&	10.6  & 1.3$ \times 10^{-8}$&	187	&	 9.8 & 2.3$ \times 10^{-9}$ \\
JI  &(2.6,4.3)		&  8.3	&	11	&	261	&	9.0	 & 3.6$ \times 10^{-8}$&	187	&	 8.3 & 3.1$ \times 10^{-9} $\\
H  &(4.2,6.1)	&	2.9	&	9.2	&	244	& 8.3	 & 4.4$ \times 10^{-8}$& 	 	&	 	 &		 \\
G  &(6.1,8.9) 	& 2.0	&	10.5	&	198	& 4.6 	 & 2.7$ \times 10^{-8}$ &		& 		 & 	 \\
F  &(8.9,10.7) 	&	2.0	&	9.8	&	176	& 3.8	 & 2.2$ \times 10^{-9}$ &		&		&	 \\
\hline
\end{tabular}
\end{minipage}
\end{table*}
The mass flux rate, \mdot\,, along the beam of the HH34 and HH1 jets has been recently measured in \cite{linda} and \cite{nisini_hh1}, respectively, using different tracers, both optical (\s\,, \oi\,) and infrared (\fe\,, H$_2$). These works show that the mass flux derived directly from the \fe\ line luminosity, using the measured tangential velocity, is always equal or larger than the \mdot\ value derived from the luminosity of the optical atomic tracers,  in spite of the possibility that part of the iron is still locked on dust grains. This is probably due to the fact that \fe\ traces a larger fraction of the total flowing mass than the optical lines, as discussed in \cite{nisini_hh1}. At the same time, it was found that in these jets the mass flux traced by the H$_2$ molecular component is negligible with respect to the mass flux due to the atomic component. Taking advantage of the velocity resolved observations  available, we now want to measure \mdot\ in the different velocity components and examine which of them is transporting more mass in the jet.


\mdot\ have been obtained here from the luminosity of the \fe\ 1.644 $\mu$m line, adopting the relationship $\dot{M} = \mu m_H \times (n_H V)\times v_t/l_t$, where $\mu = 1.24$ is the average atomic weight, $m_H$ and $n_H$ are the proton mass and the total density, $V$ is the emitting region and $v_t$ and $l_t$ are the velocity and length of the knot, projected perpendicular to the line of sight. The product $n_H V$ can be expressed as $L(line)\left(h \nu A_i f_i \frac{Fe^+}{Fe} \frac{[Fe]}{H}\right)^{-1}$. $A_i$, $f_i$ are the radiative rates and fractional population of the upper level of the considered transition and $\frac{Fe^+}{Fe}$ is the ionisation fraction of the iron having a total abundance with respect to hydrogen of $\big[\frac{Fe}{H}\big]$.

We have assumed that all iron is ionised, and has a solar abundance of $2.8\times 10^{-5}$ (\citealt{asplund05}, i.e. no dust depletion). This latter hypothesis leads to a lower limit of the actual mass flux, since it has been shown that the -velocity averaged- gas-phase iron abundance in the HH34 and HH1 jets might be only between 20 and 70\% of the solar value (\citealt{nisini_hh1,linda}). We do not adopt the Fe abundance estimate provided in these papers, since we do not know if the depletion pattern is constant among the low and high velocity components.

To compute the fractional population, we have taken the n$_e$ values derived separately for the LV and HV components, while we have assumed a single temperature for both components, equal to the values derived in \cite{nisini_hh1} for the HH1 jet and a constant value of 7000~K for HH34 given by the average value derived by \cite{linda} in knots from E to I.

Tangential velocity values have been derived from the radial velocities assuming an inclination angle i=10\degr\, for HH1 and i=22\fdg7$\pm$5\degr\, for HH34 (\citealt{bally,jochen_hh34}).  In the case of HH34, the inclination angle derived by \cite{jochen_hh34} is preferred to the value of i=30\degr\, estimated by \cite{i_hh34}, who did not consider the pattern motion of the jet knots. 
The resulting tangential velocities of the HVC derived here are listed in Table 4.  Compared  to the values  derived by the proper motion analysis of the \s\, emission by \cite{reipurth_vt_hh34}, we derive smaller velocities in the internal knots (by $\sim$ 20-50\,\kms\, in knots A6-A3) and larger in the more distant knots. In particular, we see no evidence for a deceleration of the jet as that shown by \cite{reipurth_vt_hh34}, and the velocity remains higher than 200\kms\, along the jet length. This could be an indication either that the inclination angle does not remain constant along the jet, or that the \fe\, line has a kinematical behaviour different from \s\,. 

\begin{figure}[!ht]
\begin{center}
\includegraphics[totalheight=6.5cm]{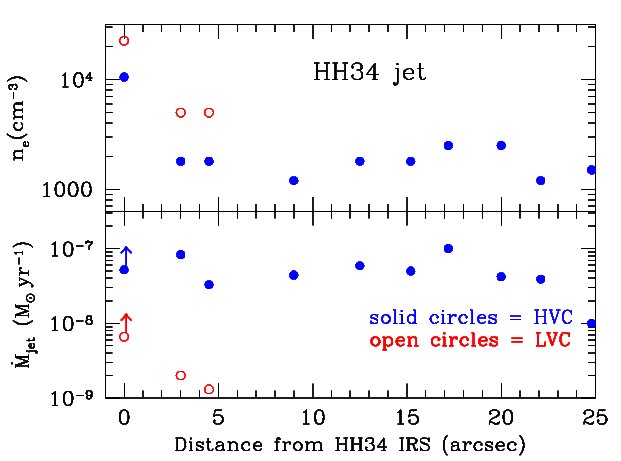}
\caption{\label{fig:mdot}
The electron density (upper panel) and mass flux (bottom panel) are represented as a function of the distance from the source for the HH34 jet. Solid circles indicate the the electron density and mass flux for the HVC, while open circles refer to the values of the LVC.}
\end{center}
\end{figure}

The luminosity of the line has been computed by integrating the extinction corrected flux of the knot in the same range of velocity used to calculate the electronic density. In Tables \ref{tab:mfluxhh34} and \ref{tab:mfluxhh1} the derived values, together with the parameters adopted, are reported for the HH34 and HH1 jets. In Figure \ref{fig:mdot}, the HH34 \mdot is plotted for the High and Low velocity components as a function of the distance from the central source.

In both jets, \mdot\,(LVC) is lower than \mdot\,(HVC) by a factor of about 6 (in HH1) and 10 (in HH34). In HH34, 
the \mdot\,(HVC)/ \mdot\,(LVC) ratio does not reflect the derived $n_e$ ratio in the two components,
indicating that  the $(n_H V)\times v_t$ product is significantly smaller in the LVC (around 8 times) than in the HVC. Since the difference in tangential velocity between each component is not enough to justify such a result, the larger \mdot\,(HVC) indicates a significant higher $(n_H V)$ product in the HVC, i.e, the HVC has either an emission volume or a total density larger than the LVC, or both. This result could be biased by the finite slit width with respect to the jet diameter. In fact, the jet width measured only by HST in the optical lines is less than 0\farcs3 up to a distance of 5\arcsec from the source \citep{reipurth_vt_hh34}. In addition, the jet emission in the low velocity component could be broader than the jet width derived from velocity integrated emission maps as observed, e.g. in DG Tau (\citealt{bacciotti00}). This effect propably does not account for the difference of nearly two orders of magnitude in derived mass-loss rates fro the LVC and HVC in the inner regions of HH34 but the derived mass-loss rates for the low-velocity component are likely lower limits. We also note that the reported values of \mdot(HVC) remain roughly constant ($\sim5\times 10^{-8}$ M$_\odot$/yr) along the whole jet as expected in steady jet flows. This fact excludes significant flux losses as the jet opening angle increases.
In jets from CTTSs, \cite{hartigan95} have already shown that  the mass flux in the LVC should be lower than in the HVC: from the analysis of not spatially resolved optical spectra, they concluded that the LVC in jets from CTT stars could be responsible for carrying the majority of mass and momentum only if the emitting region were smaller than 1 AU. 

In HH34, the mass flux derived for the internal knot A6 is very low, much smaller than the \mdot\ value in knot A3. This is likely due to the fact that we assumed here the same extinction value of 7.1 mag that was estimated by \cite{linda} as an average over the entire knot A, i.e. over 4\arcsec. The extinction value on-source has been estimated to be about 45 mag (\citealt{simone07}): therefore a large dust column density gradient is expected in the inner jet region. Assuming an A$_V$ lower limit of 7.1 mag also in the red-shifted component rA, we derive here a lower limit for \mdot\ of $\sim 1\times 10^{-9}$ M$_{\odot}$ yr$^{-1}$. The same problem probably applies to the internal knot KL in HH1, where the 8.3 mag of extinction are the average value in the whole jet region LI. 

We finally note that while the \mdot\, determined here for the HH34 jet agrees fairly well with the velocity integrated values estimated in \citet{linda}, there is some disagreement between the \mdot\, derived for HH1 here and in Nisini et al. (2005). The larger discrepancy is being found in the internal JI knots, where \cite{nisini_hh1} report a \mdot\, a factor of six larger. Part of the disagreement  is due to the different adopted tangential velocity. The largest discrepancy is, however, due to a smaller flux (a factor of three) measured in the ISAAC 0\farcs3 slit with respect to the 1\arcsec\, slit used in \cite{nisini_hh1}. It can be that this latter measurement was contaminated by the presence of the second jet responsible for the redshifted velocity component or that the HH1 jet itself has a diameter wider than 0\farcs3.

It is interesting to compare the derived mass flux rates with the source mass accretion rate, in order to estimate the  \mdot\,/$\dot{M}_{acc}$ efficiency in embedded young sources. \cite{simone07}, have derived a mass accretion rate of the order of $4.1\times10^{-6}$ M$_{\odot}$ yr$^{-1}$ for HH34 IRS. This would imply \mdot\,/$\dot{M}_{acc} \geq$ 0.01, in agreement with what is found in T Tauri stars and predicted by MHD jet launching models (e.g. \citealt{ferreira06}). There are no measurements of the mass accretion rate in VLA1, the exciting source of the HH1 jet, available. We can give a rough estimate of  this value by assuming that the source bolometric luminosity ($L_{bol} \sim$ 50 L$_{\sun}$, \citealt{chini97})  is entirely due to accretion. Such an assumption is supported by the  fact that VLA1 is a known class 0 source. If we further assume a stellar mass and radius of 1 M$_{\sun}$ and 3 R$_{\sun}$ respectively, we get $\dot{M}_{acc} \sim$\,6\,10$^{-6}$\,M$_{\odot}$\,yr$^{-1}$, which would  imply in this case \mdot\,/$\dot{M}_{acc} \geq 0.007$.

\subsection{The $Ti^+/Fe^+$ ratio in HH1}

Forbidden \Ti\ emission lines have been detected for the first time in HH1. Similarly to iron, Ti has a low ionisation potential, of only 6.82 eV, and thus is expected to be fully ionised in the jet plasma. The two detected lines have excitation energies on the order of 7000~K, and critical densities, at 10\ 000~K, $\sim5\times 10^{4}$ cm$^{-3}$. They have similar excitations, thus, with respect to \fe\ IR lines. Titanium, as iron, is a highly refractory element, thus the ratio of \Ti\ /\fe\ lines can give some clue on the relative gas-phase abundances of Ti and Fe, and, in turn on the selective depletion of these elements on dust grains still present in the jet beam. \cite{nisini_hh1} have indeed shown that the gas-phase abundance of Fe in HH1 is lower than the solar value, in particular in the inner jet regions where it is only between 20\%-30\% of the solar Fe abundance. This indicates that part of the iron is depleted on grains and consequently that a significant fraction of dust is  present in the jet beam.  We can here check if titanium follows the same depletion pattern, by comparing the observed \Ti\ 2.160 $\mu$m/ \fe\ 1.644 $\mu$m ratio with the value theoretically expected assuming solar abundance values given by \cite{asplund05}, i.e. [Fe/Ti]$_\odot$ = 354. For this analysis we have performed a statistical equilibrium calculation taking the \Ti\ radiative transition rates and electron collisional rates calculated in Bautista et al. (2006, Bautista, private communication), and considering the temperature and density values measured in each HH1 knot in \cite{nisini_hh1}. The \Ti\ 2.160 $\mu$m/\fe\ 1.644 $\mu$m intensity ratio is rather insensitive to the adopted physical conditions, being on the order of $\sim$ 500 in a temperature range from 9000 to 12\ 000 K, and density range from $5\times 10^{3}$  to $1.2\times 10^{4}$ cm$^{-3}$. The observed ratios range instead between 150 (knot JI) and 280 (knot G), implying a gas-phase Fe/Ti ratio 2-3 times lower than solar. Thus there is an overabundance of Ti in the gas phase with respect to Fe relative to the solar value. This result indicates that the release of gas-phase elements from dust grains likely follows a selective pattern in which Ti-bearing condensates are more easily 
destroyed. A large Ti/Fe abundance ratio relative to solar abundances has been observed previously in the ejecta of $\eta$ Carinae by \cite{bautista06} who suggest two different scenarios to explain this finding: either there is a spatial separation between Ti- and Fe-bearing condensates in the same dust grain, that makes the titanium more exposed to evaporation, or Ti-bearing grains are smaller than the Fe-bearing grains and thus more easily destroyed. Studies of this kind, employing a larger number of refractory species, seem therefore promising to constrain the composition  and structure of dust grains in different environments.

\section{HH34 small scale jet: comparison with models}

The \fe\ PV observed in HH34 is similar to the ones observed in some CTTSs, such as HL Tau and RW Aurigae (\citealt{pyo06}). In these latter, as in other CTTSs observed in the optical, the HVC peak has an offset with respect to the central source, of typically 50-100 AU. Given the larger distance of HH34 with respect to the sources in Taurus, we are not able to resolve such spatial regions, and the HV peak in our PV is centred on source. 
In CTT jets, the LVC is usually confined within $d \la 200$ AU from the source. In HH34, on the contrary,  the LVC at $-$60\,\kms\, persists at larger distances, up to 1000-2000 AU. In the following, we discuss these findings in the light of current theoretical models for the production of YSO jets.

Magneto-centrifugal jet launching models, such as the disc-wind and X-wind models (\citealt{ferreira97,shang98,garcia_warm,pudritz_PPV}) predict the presence of a broad velocity range close to the source position, corresponding to the uncollimated outer streamlines. In both these models, however, the LVC is confined to a relatively small region at $d \la 200$ AU from the source. Thus, the LV gas we observe at larger distance needs to be locally heated, probably by shocks.

Within 400 AU from the HH34 IRS source the line emission decreases with velocity almost continuously from its peak at \mbox{$\sim -100$\,\kms\,} to redshifted velocities up to $\sim+$200\,\kms\,.

We can compare the kinematical signatures we observe in this region with those in the synthetic PV diagrams available for the different jet models. Synthetic PV diagrams from X-wind models have been constructed for optical \s\ and \oi\ lines (\citealt{shang98}), while \fe\ PV plots have been presented by \cite{pesenti04} for the cold disc-wind model of \cite{ferreira97}.

The cold disc-wind synthetic P-V diagram presented in \cite{pesenti04} was constructed assuming a spatial sampling and spectral resolution similar to those of our observations. This diagram predicts \fe\,1.64\,\um\, maximum (de-projected) velocities on the order of  $-700$\,\kms. In contrast, the maximum velocity measured in HH34 from the FWZI in knot A6 is $\sim -$400\,\kms. However, the observed range of line of sight velocities predicted by disc-wind models significantly depends on the assumed range of launching radii and  magnetic lever arm parameter ($\lambda = (r_A/r_0)^2$, with $r_A$  and $r_0$ the the Alfv\'en and launch radii, respectively). \cite{ferreira06} have shown that cold disc-wind models in general predict too large line of sight velocities with respect to the observations, because they have too large magnetic lever arms of the order of 50 or higher. On the other hand, warm disc wind solutions with magnetic lever arms in the range 2-25 predict lower poloidal velocities, in better agreement with observations (\citealt{ferreira06}). In particular, our measured poloidal velocities  of the order of 250\,\kms\,, estimated from the observed peak radial velocity corrected for the inclination angle, would imply a value of $\lambda$ in the range 4-8, for launching radii in the range 0.07-0.15 AU  (\citealt{ferreira06}).

X-wind models predict a narrower spread in velocity that agrees with our observed range, taking  into account the inclination angle of 23\degr\, of the HH34 flow. In such models, however, the intermediate and low velocity components are predicted to be present only in the inner $\sim$ 30 AU from the central source, in contrast with observations. In addition, the synthetic PV diagrams constructed by \cite{shang98} do not include heating processes in the calculation, but assume a uniform ionisation fraction and electron temperature.

Another interesting feature predicted by X-wind models is the presence of a significant redshifted emission at the stellar position similar to what we observe in HH34. This is the first case where such a feature is observed, and indeed the absence of the redshifted emission in the PV of other CTT jets was taken as an indication to rule out the X-wind model for these sources (\citealt{takami}). In HH34, such redshifted emission was not detected in previous observations probably because of the lack of sensitivity. 

In conclusion, a qualitative comparison based on the available synthetic PV diagrams indicates that both  X-wind and disc-wind models reproduce the kinematics observed in the HH34 jet at its base, while none of them can reproduce the large  scale LVC observed at very large distances from the source. A more quantitative comparison would require the construction of \fe\, synthetic P-V diagrams exploring a larger range of parameters and heating mechanisms. 

The other observational finding in our study that can be compared to model predictions, is the dependence of the electron density and mass flux on velocity. The fact that the electron density is higher in the LVC than in the HVC is neither consistent with the available X-wind nor with disc-wind models. In disc-wind models the high velocity and dense gas ejected in the inner streamlines is bracketed by the slower gas at lower density originating from regions of the disc more external and less excited. \cite{cabrit99} show, in fact, that the electron density increases toward the axis of the flow. Similarly, in X-wind models the electron density and ionisation decreases with the distance from the central axis \citep{shu95}. 
We point out that the main parameter predicted by different jet launching models is not the electron density, but the total density and its variation with velocity. Therefore, the found inconsistency may stem from an incorrect modelling of the  excitation mechanism responsible for the gas ionisation, not necessarily from a wrong underlying MHD solution.

An alternative possibility is that the LVC represents gas not directly ejected in the jet, but dense ambient gas entrained by the high velocity collimated jet. Observations at higher spatial resolution would be needed to perform an analysis similar to that presented here for the gas within 200 AU from the source. This would allow us to isolate as much as possible the component of gas ejected at low velocity in the jet itself.

\subsection{H$_2$ high and low velocity components in HH34}

High and low velocity \h\ emission within a few hundred AU from the driving source has been detected in a dozen outflows (\citealt{davis_MHEL}); such emission regions have been called Molecular Hydrogen Emission-line region (MHEL) in analogy to the atomic FELs regions  observed in T Tauri stars. With respect to  FEL regions, the gas components associated with MHELs usually show lower radial velocities and velocity spreads. 

In light of these previous findings, there are two scenarios that can explain the characteristics of the LVC in \h\, lines in HH34. The first hypothesis is that such a low velocity warm molecular gas is excited by oblique shocks occurring at the wall of a cavity created by the interaction of a wide angle wind with the ambient medium. Indeed, the large reflection nebulosity associated with HH34 IRS suggests the presence of a cavity illuminated by the central source. The second possibility is that the \h\ gas originates from the external layers of a disc-wind. Magneto-centrifugal wind models in fact predict that the fast collimated inner component coexists with a slow and wide external component at low excitation. In a previous study, the \h\ emission observed within 100 AU from DG Tau has been interpreted by \cite{takami04} in this scenario. Modelling of the \h\ molecule survival and excitation in disc-winds is, however, needed to support this interpretation. 

The fact that the \h\ HVC is observed only from a distance of $\sim$2\arcsec\, from the star, at variance with \fe\,,      may indicate that the molecular gas does not survive in the conditions of the highly excited gas constituting the high velocity inner jet region. In this inner part  the jet is probably travelling inside a low density cavity and thus no molecular material is present to be entrained by the fast moving flow. Reformation of \h\ in the jet could be at the origin of the \h\ HVC seen further out: we point out that dust in the jet is not fully destroyed (\citealt{linda}) and thus the \h\ molecules can be formed again once the physical conditions become favourable for the molecule survival. Excitation in the wings of mini-bow shocks along the flow can be another possible explanation for the \h\ HVC, although the high radial velocity of \h\, close to the velocity of the atomic gas, is not supported by \h\, excitation models in bow-shocks (\citealt{flower03}).

\section{Conclusions}

We have presented H and K-band spectra of the HH34 and HH1 jets, where \fe\,, \h\ and for the first time \Ti\ emission lines have been detected. These observations provide us with detailed information about the kinematics of the emitting gas allowing us to resolve  two velocity components in each jet and to measure their physical parameters separately. From the \fe\ 1.600\,\um\,\,/1.644\,\um\,\ ratio the electron density along both jets and for each velocity component has been derived. In addition, the mass flux has been inferred from the \fe\ 1.644\,\um\,\ line at different positions along the jet axis and for both velocity components. Finally, the \Ti\,/\fe\, ratio gives us important information about the mechanism of dust reprocessing. Our results can be summarised as follows:
\begin{itemize}
\item{In HH34, the atomic and molecular gas show two velocity components, the so-called high (HVC) and low (LVC) velocity components, near to the source. We also detect for the first time the fainter red-shifted counterflow down to the central source and up to a distance of 25\arcsec\, from the star.}
\item{HH1 is traced down to $\sim 1\arcsec$ from the source. The kinematics of the region closest to the driving source is again characterised by a double velocity component, one blue-shifted and one red-shifted with respect to the source LSR. We interpret the red-shifted component as part of another jet, previously detected by \cite{Reipurth_hh1} in optical and near-IR lines.}
\item{In the innermost HH34 jet region, the electron density increases as the velocity of the jet decreases, with average values of $10.8\times 10^3$ cm$^{-3}$ and $4.7\times 10^3$ cm$^{-3}$ for the LV and HV components, respectively. On the contrary, at large distance from the source, in HH34 as well as in HH1, $n_e$ increases with velocity. The average values of the electron density along HH1 are  $n_e\sim 9.8\times 10^3$ cm$^{-3}$ and $n_e\sim 9\times 10^{3}$ cm$^{-3}$ for the blue- and red-shifted components, respectively.
}

\item{The mass flux is mainly carried by the high velocity gas in both jets. We derive a lower limit on the mass flux of $3-8\times 10^{-8}$ M$_{\odot}$ yr$^{-1}$ along the HH34 and HH1 jets.}

\item{Comparing the observed \Ti\,/\fe\ ratio with the theoretical one, we derived a gas-phase Fe/Ti abundance ratio 2-3 times less than solar. This seems to indicate that the release of gas-phase elements from dust grains likely follows a selective pattern in which Ti-bearing condensates are more easily destroyed than the Fe ones.}

\item{By a comparison of our PV diagrams and electron densities with those provided by models for MHD jet launching, we suggest that the kinematical features observed close to the source in our spectra can be, qualitatively, reproduced by both disc-wind and X-wind models, although none of them is able to explain the persistency of the LVC at large distances (up to 1000 AU) from the launching region. Moreover, none of the excitation mechanisms proposed in these models  can explain the dependence of electron density with velocity that we measure in HH34. Alternatively, the LVC of the \fe\, lines that we observe in HH34 could represent dense gas entrained by the high velocity collimated jet.}

\item{In the \h\ line, only the LVC is observed down to the central source, while the HVC is detected only up to 2\arcsec\, from the star. We suggest that the low velocity molecular gas could be excited by oblique shocks occurring along the wall of a cavity created by the interaction of a wide angle wind or by the external layers of a disc-wind. On the other hand, the high velocity component of the \h\, emission could be due to reformation of \h\ molecules along the jet, since the dust in the jet is probably not fully destroyed. }

\end{itemize}

\begin{acknowledgements}
The present work was partly supported by the European Community’s
Marie Curie Actions - Human Resource and Mobility within the JETSET (Jet
Simulations, Experiments and Theory) network under contract MRTN-CT-2004
005592. We thank Manuel Bautista for having provided us with the atomic data for the \Ti\, 
statistical equilibrium model.
\end{acknowledgements}

\bibliographystyle{aa}
\bibliography{references}

\begin{thebibliography}{49}
\expandafter\ifx\csname natexlab\endcsname\relax\def\natexlab#1{#1}\fi

\bibitem[{{Anglada} {et~al.}(1995){Anglada}, {Estalella}, {Mauersberger},
  {Torrelles}, {Rodriguez}, {Canto}, {Ho}, \& {D'Alessio}}]{v_cloud_hh34}
{Anglada}, G., {Estalella}, R., {Mauersberger}, R., {et~al.} 1995, \apj, 443,
  682

\bibitem[{{Antoniucci} {et~al.}(2008){Antoniucci}, {Nisini}, {Giannini}, \&
  {Lorenzetti}}]{simone07}
{Antoniucci}, S., {Nisini}, B., {Giannini}, T., \& {Lorenzetti}, D. 2008, \aap,
  479, 503

\bibitem[{{Asplund} {et~al.}(2005){Asplund}, {Grevesse}, \&
  {Sauval}}]{asplund05}
{Asplund}, M., {Grevesse}, N., \& {Sauval}, A.~J. 2005, in Astronomical Society
  of the Pacific Conference Series, Vol. 336, Cosmic Abundances as Records of
  Stellar Evolution and Nucleosynthesis, ed. T.~G. {Barnes}, III \& F.~N.
  {Bash}, 25--+

\bibitem[{{Bacciotti} {et~al.}(2000){Bacciotti}, {Mundt}, {Ray},
  {Eisl{\"o}ffel}, {Solf}, \& {Camezind}}]{bacciotti00}
{Bacciotti}, F., {Mundt}, R., {Ray}, T.~P., {et~al.} 2000, \apjl, 537, L49

\bibitem[{{Bally} {et~al.}(2002){Bally}, {Heathcote}, {Reipurth}, {Morse},
  {Hartigan}, \& {Schwartz}}]{bally}
{Bally}, J., {Heathcote}, S., {Reipurth}, B., {et~al.} 2002, \aj, 123, 2627

\bibitem[{{Bautista} {et~al.}(2006){Bautista}, {Hartman}, {Gull}, {Smith}, \&
  {Lodders}}]{bautista06}
{Bautista}, M.~A., {Hartman}, H., {Gull}, T.~R., {Smith}, N., \& {Lodders}, K.
  2006, \mnras, 370, 1991

\bibitem[{{Beck} {et~al.}(2007){Beck}, {Riera}, {Raga}, \& {Reipurth}}]{beck07}
{Beck}, T.~L., {Riera}, A., {Raga}, A.~C., \& {Reipurth}, B. 2007, \aj, 133,
  1221

\bibitem[{{Cabrit} {et~al.}(1999){Cabrit}, {Ferreira}, \& {Raga}}]{cabrit99}
{Cabrit}, S., {Ferreira}, J., \& {Raga}, A.~C. 1999, \aap, 343, L61

\bibitem[{{Camenzind}(1990)}]{camenzind90}
{Camenzind}, M. 1990, in Reviews in Modern Astronomy, Vol.~3, Reviews in Modern
  Astronomy, ed. G.~{Klare}, 234--265

\bibitem[{{Chini} {et~al.}(1997){Chini}, {Reipurth}, {Sievers},
  {Ward-Thompson}, {Haslam}, {Kreysa}, \& {Lemke}}]{chini97}
{Chini}, R., {Reipurth}, B., {Sievers}, A., {et~al.} 1997, \aap, 325, 542

\bibitem[{{Chini} {et~al.}(2001){Chini}, {Ward-Thompson}, {Kirk}, {Nielbock},
  {Reipurth}, \& {Sievers}}]{chini01}
{Chini}, R., {Ward-Thompson}, D., {Kirk}, J.~M., {et~al.} 2001, \aap, 369, 155

\bibitem[{{Choi} \& {Zhou}(1997)}]{v_cloud_hh1}
{Choi}, M. \& {Zhou}, S. 1997, \apj, 477, 754

\bibitem[{{Davis} {et~al.}(2006){Davis}, {Nisini}, {Takami}, {Pyo}, {Smith},
  {Whelan}, {Ray}, \& {Chrysostomou}}]{davisSVS13}
{Davis}, C.~J., {Nisini}, B., {Takami}, M., {et~al.} 2006, \apj, 639, 969

\bibitem[{{Davis} {et~al.}(2001){Davis}, {Ray}, {Desroches}, \&
  {Aspin}}]{davis_MHEL}
{Davis}, C.~J., {Ray}, T.~P., {Desroches}, L., \& {Aspin}, C. 2001, \mnras,
  326, 524

\bibitem[{{Davis} {et~al.}(2000){Davis}, {Smith}, \& {Eisl{\"o}ffel}}]{davis00}
{Davis}, C.~J., {Smith}, M.~D., \& {Eisl{\"o}ffel}, J. 2000, \mnras, 318, 747

\bibitem[{{Davis} {et~al.}(2003){Davis}, {Whelan}, {Ray}, \&
  {Chrysostomou}}]{davis03}
{Davis}, C.~J., {Whelan}, E., {Ray}, T.~P., \& {Chrysostomou}, A. 2003, \aap,
  397, 693

\bibitem[{{Eisl{\"o}ffel} \& {Mundt}(1992)}]{jochen_hh34}
{Eisl{\"o}ffel}, J. \& {Mundt}, R. 1992, \aap, 263, 292

\bibitem[{{Eisl{\"o}ffel} {et~al.}(1994){Eisl{\"o}ffel}, {Mundt}, \&
  {B{\"o}hm}}]{jochen94}
{Eisl{\"o}ffel}, J., {Mundt}, R., \& {B{\"o}hm}, K.-H. 1994, \aj, 108, 1042

\bibitem[{{Eisl{\"o}ffel} {et~al.}(2000){Eisl{\"o}ffel}, {Smith}, \&
  {Davis}}]{jochen00}
{Eisl{\"o}ffel}, J., {Smith}, M.~D., \& {Davis}, C.~J. 2000, \aap, 359, 1147

\bibitem[{{Ferreira}(1997)}]{ferreira97}
{Ferreira}, J. 1997, \aap, 319, 340

\bibitem[{{Ferreira} {et~al.}(2006){Ferreira}, {Dougados}, \&
  {Cabrit}}]{ferreira06}
{Ferreira}, J., {Dougados}, C., \& {Cabrit}, S. 2006, \aap, 453, 785

\bibitem[{{Flower} {et~al.}(2003){Flower}, {Le Bourlot}, {Pineau des
  For{\^e}ts}, \& {Cabrit}}]{flower03}
{Flower}, D.~R., {Le Bourlot}, J., {Pineau des For{\^e}ts}, G., \& {Cabrit}, S.
  2003, \mnras, 341, 70

\bibitem[{{Garcia} {et~al.}(2001){Garcia}, {Ferreira}, {Cabrit}, \&
  {Binette}}]{garcia_warm}
{Garcia}, P.~J.~V., {Ferreira}, J., {Cabrit}, S., \& {Binette}, L. 2001, \aap,
  377, 589

\bibitem[{{Hamann} {et~al.}(1994){Hamann}, {Simon}, {Carr}, \&
  {Prato}}]{hamann94}
{Hamann}, F., {Simon}, M., {Carr}, J.~S., \& {Prato}, L. 1994, \apj, 436, 292

\bibitem[{{Hartigan} {et~al.}(1995){Hartigan}, {Edwards}, \&
  {Ghandour}}]{hartigan95}
{Hartigan}, P., {Edwards}, S., \& {Ghandour}, L. 1995, \apj, 452, 736

\bibitem[{{Hartigan} {et~al.}(1994){Hartigan}, {Morse}, \&
  {Raymond}}]{hartigan94}
{Hartigan}, P., {Morse}, J.~A., \& {Raymond}, J. 1994, \apj, 436, 125

\bibitem[{{Hartigan} {et~al.}(1987){Hartigan}, {Raymond}, \&
  {Hartmann}}]{hartigan87}
{Hartigan}, P., {Raymond}, J., \& {Hartmann}, L. 1987, \apj, 316, 323

\bibitem[{{Heathcote} \& {Reipurth}(1992)}]{i_hh34}
{Heathcote}, S. \& {Reipurth}, B. 1992, \aj, 104, 2193

\bibitem[{{Hirth} {et~al.}(1997){Hirth}, {Mundt}, \& {Solf}}]{hirth97}
{Hirth}, G.~A., {Mundt}, R., \& {Solf}, J. 1997, \aaps, 126, 437

\bibitem[{{Lavalley-Fouquet} {et~al.}(2000){Lavalley-Fouquet}, {Cabrit}, \&
  {Dougados}}]{lavalley00}
{Lavalley-Fouquet}, C., {Cabrit}, S., \& {Dougados}, C. 2000, \aap, 356, L41

\bibitem[{{Matt} \& {Pudritz}(2005)}]{stellar_wind05}
{Matt}, S. \& {Pudritz}, R.~E. 2005, \apjl, 632, L135

\bibitem[{{Nisini} {et~al.}(2005){Nisini}, {Bacciotti}, {Giannini}, {Massi},
  {Eisl{\"o}ffel}, {Podio}, \& {Ray}}]{nisini_hh1}
{Nisini}, B., {Bacciotti}, F., {Giannini}, T., {et~al.} 2005, \aap, 441, 159
  (N05)

\bibitem[{{Nisini} {et~al.}(2002){Nisini}, {Caratti o Garatti}, {Giannini}, \&
  {Lorenzetti}}]{nisini02}
{Nisini}, B., {Caratti o Garatti}, A., {Giannini}, T., \& {Lorenzetti}, D.
  2002, \aap, 393, 1035

\bibitem[{{Pesenti} {et~al.}(2003){Pesenti}, {Dougados}, {Cabrit}, {O'Brien},
  {Garcia}, \& {Ferreira}}]{pesenti04}
{Pesenti}, N., {Dougados}, C., {Cabrit}, S., {et~al.} 2003, \aap, 410, 155

\bibitem[{{Podio} {et~al.}(2006){Podio}, {Bacciotti}, {Nisini},
  {Eisl{\"o}ffel}, {Massi}, {Giannini}, \& {Ray}}]{linda}
{Podio}, L., {Bacciotti}, F., {Nisini}, B., {et~al.} 2006, \aap, 456, 189 (P06)

\bibitem[{{Pudritz} {et~al.}(2007){Pudritz}, {Ouyed}, {Fendt}, \&
  {Brandenburg}}]{pudritz_PPV}
{Pudritz}, R.~E., {Ouyed}, R., {Fendt}, C., \& {Brandenburg}, A. 2007, in
  Protostars and Planets V, ed. B.~{Reipurth}, D.~{Jewitt}, \& K.~{Keil},
  277--294

\bibitem[{{Pyo} {et~al.}(2002){Pyo}, {Hayashi}, {Kobayashi}, {Terada}, {Goto},
  {Yamashita}, {Tokunaga}, \& {Itoh}}]{pyo02}
{Pyo}, T.-S., {Hayashi}, M., {Kobayashi}, N., {et~al.} 2002, \apj, 570, 724

\bibitem[{{Pyo} {et~al.}(2006){Pyo}, {Hayashi}, {Kobayashi}, {Tokunaga},
  {Terada}, {Takami}, {Takato}, {Davis}, {Takami}, {Hayashi}, {G{\"a}ssler},
  {Oya}, {Hayano}, {Kamata}, {Minowa}, {Iye}, {Usuda}, {Nishikawa}, \&
  {Nedachi}}]{pyo06}
{Pyo}, T.-S., {Hayashi}, M., {Kobayashi}, N., {et~al.} 2006, \apj, 649, 836

\bibitem[{{Pyo} {et~al.}(2003){Pyo}, {Kobayashi}, {Hayashi}, {Terada}, {Goto},
  {Takami}, {Takato}, {Gaessler}, {Usuda}, {Yamashita}, {Tokunaga}, {Hayano},
  {Kamata}, {Iye}, \& {Minowa}}]{pyo03}
{Pyo}, T.-S., {Kobayashi}, N., {Hayashi}, M., {et~al.} 2003, \apj, 590, 340

\bibitem[{{Raga} \& {Noriega-Crespo}(1998)}]{raga98}
{Raga}, A. \& {Noriega-Crespo}, A. 1998, \aj, 116, 2943

\bibitem[{{Raga} {et~al.}(2002){Raga}, {Vel{\'a}zquez}, {Cant{\'o}}, \&
  {Masciadri}}]{raga02}
{Raga}, A.~C., {Vel{\'a}zquez}, P.~F., {Cant{\'o}}, J., \& {Masciadri}, E.
  2002, \aap, 395, 647

\bibitem[{{Reipurth} {et~al.}(2002){Reipurth}, {Heathcote}, {Morse},
  {Hartigan}, \& {Bally}}]{reipurth_vt_hh34}
{Reipurth}, B., {Heathcote}, S., {Morse}, J., {Hartigan}, P., \& {Bally}, J.
  2002, \aj, 123, 362

\bibitem[{{Reipurth} {et~al.}(2000){Reipurth}, {Heathcote}, {Yu}, {Bally}, \&
  {Rodr{\'{\i}}guez}}]{Reipurth_hh1}
{Reipurth}, B., {Heathcote}, S., {Yu}, K.~C., {Bally}, J., \&
  {Rodr{\'{\i}}guez}, L.~F. 2000, \apj, 534, 317

\bibitem[{{Shang} {et~al.}(2002){Shang}, {Glassgold}, {Shu}, \&
  {Lizano}}]{shang02}
{Shang}, H., {Glassgold}, A.~E., {Shu}, F.~H., \& {Lizano}, S. 2002, \apj, 564,
  853

\bibitem[{{Shang} {et~al.}(1998){Shang}, {Shu}, \& {Glassgold}}]{shang98}
{Shang}, H., {Shu}, F.~H., \& {Glassgold}, A.~E. 1998, \apjl, 493, L91+

\bibitem[{{Shu} {et~al.}(1995){Shu}, {Najita}, {Ostriker}, \& {Shang}}]{shu95}
{Shu}, F.~H., {Najita}, J., {Ostriker}, E.~C., \& {Shang}, H. 1995, \apjl, 455,
  L155+

\bibitem[{{Solf} {et~al.}(1991){Solf}, {Raga}, {Boehm}, \&
  {Noriega-Crespo}}]{solf91}
{Solf}, J., {Raga}, A.~C., {Boehm}, K.~H., \& {Noriega-Crespo}, A. 1991, \aj,
  102, 1147

\bibitem[{{Takami} {et~al.}(2004){Takami}, {Chrysostomou}, {Ray}, {Davis},
  {Dent}, {Bailey}, {Tamura}, \& {Terada}}]{takami04}
{Takami}, M., {Chrysostomou}, A., {Ray}, T.~P., {et~al.} 2004, \aap, 416, 213

\bibitem[{{Takami} {et~al.}(2006){Takami}, {Chrysostomou}, {Ray}, {Davis},
  {Dent}, {Bailey}, {Tamura}, {Terada}, \& {Pyo}}]{takami}
{Takami}, M., {Chrysostomou}, A., {Ray}, T.~P., {et~al.} 2006, \apj, 641, 357

\end{thebibliography}

\end{document}